\documentclass[twocolumn, pra, aps, amsfonts, amsmath, amssymb, showkeys]{revtex4-1} 

\usepackage{graphicx,bbm,enumerate,float,xcolor}
\usepackage[colorlinks=true,bookmarks=false,citecolor=blue,urlcolor=blue]{hyperref}
\usepackage{caption}
\usepackage{subcaption}
\graphicspath{ {./images/} }
\usepackage{tikz}
\usetikzlibrary{quantikz2}
\usepackage[utf8]{inputenc}
\usepackage[export]{adjustbox}
\usepackage{physics}
\usepackage{comment}
\usepackage[shortlabels]{enumitem}
\everymath{\displaystyle}

\newcommand{\orcid}[1]{\href{https://orcid.org/#1}{\includegraphics[width=10pt]{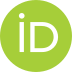}}}

\newcommand{\beq}{\begin{align}}
\newcommand{\eeq}{\end{align}}
\begin{document}

\title{A Sharp Geometric Measure of Entanglement}

\author{Dharmaraj Ramachandran\orcid{0000-0002-3068-1586}}
\email{p20200040@goa.bits-pilani.ac.in}

\author{Radhika Vathsan\orcid{0000-0001-5892-9275}}
\email{radhika@goa.bits-pilani.ac.in}

\affiliation{Physics Department,\\ BITS-Pilani, K. K. Birla Goa Campus, Goa 403726, INDIA}

\date{\today}

\begin{abstract}
Despite their elegance and widespread use, the current Geometric Measures (GMs) of entanglement exhibit a significant limitation: they fail to effectively distinguish Local Unitary (LU) inequivalent states due to the inherent nature of their definition. We illustrate the impact of this limitation using the fidelity of the teleportation protocol as an example. To address this issue, we introduce the Sharp Geometric Measure (SGM) by modifying the standard definition of the Geometric Measure. We show that the closed-form expression of the SGM can be equivalently derived using the Riemannian structure of both the composite state space and the reduced density operator space. Furthermore, we define a measure of Genuine Multipartite Entanglement (GME) derived from the SGM, which we term GMS. We demonstrate that GMS resolves two key limitations of some existing GME measures, thereby establishing its utility and effectiveness in quantifying GME.
\end{abstract}

\maketitle

\section{Introduction}

Entanglement has become a  signature property of quantum systems that is an important resource in quantum computing and communication. Quantification of entanglement in different contexts is therefore a subject of intense research in current times.

 Historically, several measures of pure state entanglement have been formulated, exploring different aspects of what appears to be a multi-faceted property. Algebraic measures such as   concurrence~\cite{concurrence_1998,concurrence_2}, negativity~\cite{Eisert1999} and reshuffling negativity~\cite{Chen2002}  are  straightforward and  have been proven to be computationally inexpensive. There are operational measures such as  entanglement cost~\cite{Bennett1996} and distillable entanglement~\cite{dist_ent_1} based on the amount of resources required to generate entanglement. Among different approaches there have  been measures of entanglement based on operator algebra~\cite{Balachandran2013} and more recently,  coherence of an entangled state in the Schmidt basis~\cite{Pathania_2022}.

Geometric entanglement measures are formulated in terms of  the distance between the state under consideration and the closest  separable state. Several authors~\cite{ent_eigval_Gobart,Vedral_1998,Eisert_2003,Witte_1999} have approached such measures  using different metrics on quantum state space.
Amari and Nagaoka~\cite{amari2000methods} provided a detailed study of quantum information based on the differential geometric structure of quantum state space, which has also been studied in  great depth by Provost et al,\cite{Provost1980}, Morozova and \c{C}entsov~\cite{Morozova1989} and  Petz \cite{Petz1996}.

The necessary properties of an entanglement measure for bipartite systems are~\cite{Zyczkowski2006}:
\textbf{(E1)}~Monotonocity under local operations and classical communications (LOCC), \textbf{(E2)}~Discriminance: $E(\rho) = 0$ iff $\rho$ is separable, \textbf{(E3)}~Convexity, \textbf{(E4)}\label{E4}~Normalizability and 
\textbf{(E5)}\label{E5}~Computability.
 

  Defining entanglement measures for  multipartite systems is complicated due to the existence of different {\it kinds} of entanglement.  
 It has been proven that Genuine Multipartite Entanglement(GME) is a fundamental resource in various quantum information processing tasks~\cite{GME_resource_1,GME_resource_2,GME_resource_3,GME_resource_4,GME_resource_5,GME_resource_6}.
 A multipartite entanglement monotone $E(\rho)$ is a measure of GME~\cite{GME_intro1,GME_intro2} if it satisfies the following properties (\cite{Gme_props_Ma_2011,GmE_props_2,Choi2023_proper, Joo_2003_proper}):
\textbf{(P1)}
    \label{P1} $E(\rho)=0$ for all product states {\em and} for all biseparable states,  
   \label{P2}\textbf{(P2)} $E(\rho)> 0$ for all non-biseparable states.   

Some existing multipartite entanglement measures seem to fall short of one or other of these conditions. For example, 
 the multi-qubit entanglement measure given by Meyer \textit{et al}~\cite{Global_ent_1} and interpreted by  Brennen~\cite{globalent_2}  is non-zero for biseparable states. Later Love \textit{et al}\cite{Love2007} gave a correction to this  measure which qualified it  as a measure of  GME. In works by Li \textit{et al }\cite{GM_of_conc}, Xi \textit{et al}~\cite{schmidt_gme}, Sab{\'i}n \textit{et al}~\cite{Classification_ent2008} and Markiewicz \textit{et al}~\cite{gme_bip_correlations}, bipartite measures have been utilised to construct a measure of GME. The measures given by Pan \textit{et al}~\cite{sumofs_Pan2004} and by Carvalho \textit{et al}~\cite{concent_ent} use addition of bipartite measures, which in general are non-zero for biseparable states and thus do  not satisfy \textbf{(P1)}. Cai \textit{et al}~\cite{info_theoritic_gme} have defined a GME  based on von Neumann entropy. The well-known measure 3-tangle, which is an algebraic extension of concurrence given by Coffman \textit{et al}~\cite{tangle_1} and later extended by Miyake~\cite{ tangle_2},  does not satisfy \textbf{(P2)} and is always zero for W-class entangled three qubit states\eqref{W}. Bounds on geometric measures of entanglement  using z-spectral radius for some classes of pure states were given by Xiong \textit{et al}~\cite{zeigen_val_gme_Xiong2022}. 
Haddadi \textit{et al}~\cite{review_of_geometric_gme}  and Mengru Ma \textit{et al}~\cite{gme_review} give  an overview of measures of multi-qubit entanglement.

In this work, we construct a new geometric measure  of entanglement that addresses an issue of  concern with geometric entanglement measures in general: a notion we call ``sharpness". We highlight the relevance of this issue using quantum teleportation as an illustrative example. We call our modified   geometric measure  the Sharp Geometric Measure (SGM) and derive a closed-form expression for pure bipartite states. Building on SGM, we define a measure of Genuine Multipartite Entanglement (GME), which we term the Geometric Mean of the Sharp Geometric Measure (GMS). Through illustrative examples, we demonstrate how both SGM and GMS overcome some of the limitations of existing entanglement measures.

\section{A shortcoming of  geometric measures of entanglement}
The standard definition of a geometric measure of entanglement for a bipartite state $\ket{\psi} \in \mathcal{H}_A \otimes \mathcal{H}_B$ is the distance to the closest separable state: 
\begin{equation}
    GM(\psi) = \min_{\phi_s \in \mathcal{S}_{\text{sep}} }D(\psi , \phi_s),
    \label{eq:closest_sep}
\end{equation} 
where $\mathcal{S}_{\text{sep}}$ denotes the set of all separable states, and $D(.,.)$ is a distance measure on the space of states.

This notion of distance to the closest separable state has been the foundation for several entanglement measures \cite{ent_eigval_Gobart,Vedral_1998,Eisert_2003,Witte_1999} in the literature. While there exist various geometric measures, they all differ only in the choice of the distance measure used. Though definition poses the problem of minimization over an infinite set, the Schmidt decomposition facilitates finding a closed-form solution for Eq.~\eqref{eq:closest_sep} in the case of pure bipartite states. 

The Schmidt decomposition of a pure bipartite state is expressed as 
\begin{equation}
    \ket{\psi} = \sum_i \sqrt{\lambda_i}\ket{a_i} \otimes \ket{b_i},
    \label{eq:schm_decomp}
\end{equation}
where $\ket{a_i}$ and $\ket{b_i}$ are the eigenvectors of the  density operators of subsystems $A$ and $B$ respectively,  $\lambda_i \in [0 , 1]$ being their corresponding eigenvalues. 

Wei \textit{et al.} \cite{ent_eigval_Gobart} demonstrated that the closest separable state to a pure bipartite state is  given by the Schmidt vector corresponding to the largest Schmidt coefficient $\lambda_{\text{max}}$, 
which they  call  the  entanglement eigenvalue. 
All geometric measures, such as those in \cite{Vedral_1998,Eisert_2003,Witte_1999}, are functions of $\lambda_{\text{max}}$. Thus Geometric Measures of entanglement (GMs) depend solely on the largest Schmidt coefficient. This  implies that GM's are insensitive to variations in the non-maximal Schmidt coefficients.

However,  all the Schmidt coefficients play a role in  measuring the entanglement content of a state.
If the Schmidt coefficients of two pure bipartite states $\ket{\psi_1}$ and $\ket{\psi_2}$ are different, then  the unitary transformation $U_{AB}$ relating them is by definition non-local:
\begin{align*}
    U_{AB}\ket{\psi_1} = \ket{\psi_2} \implies U_{AB} \neq U_A \otimes U_{B}.
\end{align*}
The entanglement content of a state is preserved only under local unitary (LU) transformations.  GM's are  therefore  insensitive to such transformations that may not  preserve the entanglement content of the state. 
We would like to call  measures that are sensitive to changes in any Schmidt coefficient as \textit{sharp} entanglement measures. 
Certain  measures like entanglement entropy~\cite{ent_entropy} are sharp by definition while GM's are not. In the next section we highlight the importance of sharpness in entanglement quantification using quantum teleportation as an example.

\section{Quantum Teleportation: A Case for Sharpness}
This example, we demonstrate that the success of a teleportation protocol, as measured by the average fidelity ($\mathcal{F}_{\text{av}}$), depends on non-maximal Schmidt coefficients. This study naturally necessitates quantum states with more than two Schmidt coefficients. Consequently, we have chosen the simplest case: a two-qutrit state with three Schmidt coefficients.

\noindent \textbf{Setup:} Alice intends to teleport  to Bob an arbitrary single-qubit state 
\begin{equation*}
    \ket{\psi(\alpha,\theta)} = \sqrt{\alpha}\ket{0} + \sqrt{1-\alpha} e^{i\theta}\ket{1},
\end{equation*}
where $\alpha \in [0, 1]$ and $\theta \in [0, 2\pi]$. Alice and Bob share a maximally entangled two-qutrit state given by 
\begin{equation}
    \ket{\beta}_{AB} = \frac{1}{\sqrt{3}} (\ket{00} + \ket{11} + \ket{22}).
\end{equation}

To begin the protocol, Alice performs a measurement in the basis $\{\ket{\mu}\}$  of  maximally entangled (Bell-like) qubit-qutrit states, which in the computational basis are
\begin{align*}
  \{\ket{\mu}\}=   \left\{ \frac{\ket{00} \pm \ket{11}}{\sqrt{2}}, \frac{\ket{01} \pm \ket{12}}{\sqrt{2}}, 
    \frac{\ket{02} \pm \ket{10}}{\sqrt{2}} \right\}.
\end{align*}

\begin{figure}
\centering
\begin{quantikz}[row sep=0.1cm]
\lstick{$\ket{\psi(\alpha,\phi)}$}&\gate[2][2cm]{\parbox{2cm}{$\ket{\mu}$-basis measurement}}&\wire{c}&\ctrl[vertical wire=c]{2}\wire{c} \\
\lstick[2]{$\ket{\beta}$}&&\ctrl[vertical wire=c]{1}\wire{c} \\
&&\gate{U_2}&\gate{U_1}&\rstick{$\rho_{\text{out}}$}
\end{quantikz}
\caption{Single-qubit teleportation using a two-qutrit  maximally entangled state.}
\label{fig:teleport}
\end{figure}
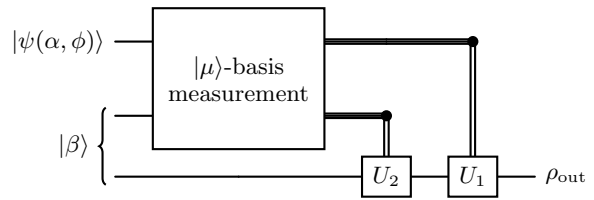

After the measurement, Alice conveys the outcome to Bob via a classical communication channel. Based on the measurement outcomes, Bob applies appropriate unitary gates as shown in Fig.~\ref{fig:teleport} to retrieve the state $\ket{\psi(\alpha,\theta)}$. 
These gates conditioned on Alice's  measurement outcome are defined as the  following controlled gates: 
\begin{align*}
    CU_{1} &= \op{0}{0} \otimes \mathbbm{I}_3 + \op{1}{1} \otimes U_{a}, \\
    CU_{2} &= \op{0}{0} \otimes \mathbbm{I}_3 + \op{1}{1} \otimes U_{b} + \op{2}{2} \otimes U_c,
\end{align*}
where the $U_i$ operators are unitary transformations expressed in the computational qutrit basis as:
\begin{align*}
    U_a &= 
    \begin{bmatrix}
    1 & 0 & 0 \\ 
    0 & -1 & 0 \\ 
    0 & 0 & 1
    \end{bmatrix}, \quad
    U_b = 
    \begin{bmatrix}
    0 & 1 & 0 \\ 
    0 & 0 & 1 \\ 
    1 & 0 & 0
    \end{bmatrix}, \quad
    U_c =
    \begin{bmatrix}
    0 & 0 & 1 \\ 
    1 & 0 & 0 \\ 
    0 & 1 & 0
    \end{bmatrix}.
\end{align*}

The fidelity between the input state $\ket{\psi(\alpha,\theta)}$ and the output state $\rho_{\text{out}}$ is given by~\cite{Bures_orginal_paper} 
\begin{equation*}
    \mathcal{F}(\psi(\alpha,\theta), \rho_{\text{out}}) = \sqrt{\bra{\psi(\alpha,\theta)}\rho_{\text{out}}\ket{\psi(\alpha,\theta)}}.
\end{equation*}
The success of the protocol is determined by the fidelity averaged ($\mathcal{F}_{\text{av}}$) over all values of $\alpha$ and $\theta$. 

\begin{figure}[h!]
\centering
\includegraphics[width=0.4\textwidth]{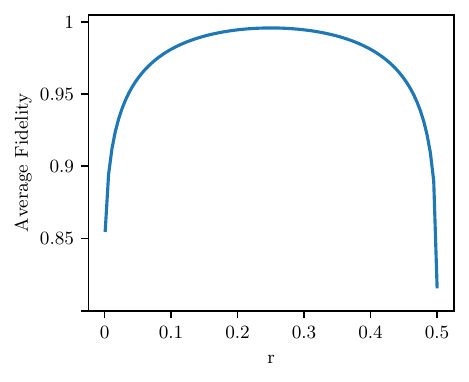}
\caption{Average fidelity of teleportation as a function of $r$ of $\ket{\beta(r)}$ in Eq: \eqref{eq:beta_r}.}
\label{fig:fid_av}
\end{figure}

When the shared entangled pair between Alice and Bob is $\ket{\beta}$, it is straightforward to see that 
\begin{equation*}
    \mathcal{F} = 1, \; \forall \;\alpha,\;\phi \implies \mathcal{F}_{\text{av}} = 1.
\end{equation*}
However, when the shared entangled pair is not a maximally entangled state but instead 
\begin{equation}
    \ket{\beta(r)} = \sqrt{0.5} \ket{00} + \sqrt{0.5 - r} \ket{11} + \sqrt{r} \ket{22},
    \label{eq:beta_r}
\end{equation}
where $r \in [0, 0.5]$, the average fidelity $\mathcal{F}_{\text{av}}$ is not  constant, but varies with $r$ as shown in Fig.~\ref{fig:fid_av}. 

The dependence on $r$ of teleportation fidelity indicates a practical outcome of the difference in  entanglement content of $\ket{\beta(r)}$ for different $r$. This difference is not picked up by geometric measures (GM): 
\begin{equation*}
    GM(\beta(r)) = 0.75 \quad \forall \; r \in [0, 0.5].
\end{equation*}
While the fidelity of the teleportation protocol indicates a change in the entanglement content for states like $\ket{\beta(r)}$, standard geometric measures fail to capture this variation. This limitation arises because standard geometric measures are insensitive to changes in non-maximal Schmidt coefficients. 

To address this limitation, we introduce a new entanglement measure, the Sharp Geometric Measure (SGM), that has contribution from all Schmidt coefficients.

\section{Sharp Geometric Measure (SGM)}
We propose a modification to the Geometric Measure of Entanglement in order to make it sharp. The Sharp Geometric Measure (SGM) is defined as the complement of the normalized distance to the closest maximally entangled state.  For a bipartite pure state $\ket{\psi}$, SGM is expressed as:
\begin{equation}
    \text{SGM}(\psi) = 1 - \min_{\Psi_m \in \mathcal{S}_{\text{max}}}\frac{D_{F}(\psi, \Psi_m)}{N_1} ,
    \label{eq:closest_max}
\end{equation}
where $\mathcal{S}_{\text{max}}$ is the set of all maximally entangled states, $D_{F}(.,.)$ denotes the Fubini-Study distance between quantum states, and $N_1$ is a normalization constant ensuring that $D_F(\psi,\Psi_m)/N_1 \in [0, 1]$. The normalization constant will become evident once a closed-form expression for Eq.~\eqref{eq:closest_max} is derived. We choose the Fubini-Study distance due to its status as the only Riemannian distance defined in the space of pure states.

The notion of distance to closest maximally entangled state was introduced earlier as maximal fidelity~\cite{mf1,mf_2,mf_3}. Its utility as a measure of entanglement is less pursued as it was pointed out by K. \.Zyczkowski \textit{et al}~\cite{Bengtsson2006} that maximal fidelity does not satisfy \textbf{E2} and it was shown by P.Badziag \textit{et al}~\cite{horo_mf_locc} and Verstraete \textit{et al}~\cite{mf_3} that it also violates \textbf{E1}. While maximal fidelity is not an entanglement measure, SGM does satisfy necessary conditions for a valid entanglement measure which will be explained after we derive the closed form expression. 

\textbf{Claim:} The  maximally entangled state $\ket{\Psi_c}$ closest to a bipartite pure state $\ket{\psi} \in \mathcal{H}_A \otimes \mathcal{H}_B$ is the one with the same Schmidt vectors as $\ket{\psi}$, with all Schmidt coefficients equal.

\begin{figure}[h]
\centering
\includegraphics[width=0.38\textwidth]{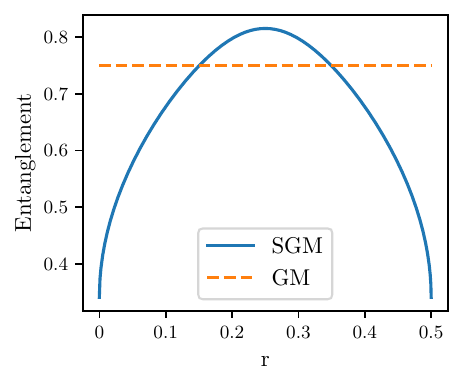} 
\caption{Comparison of SGM and GM for $\ket{\beta(r)}$. }
\label{LU_ineq}
\end{figure}

\textbf{Proof:} Let the Schmidt representation of  $\ket{\psi}$ be
\begin{equation}
    \ket{\psi} = \sum_{i=0}^{n_A - 1} \sqrt{\lambda_i} \ket{a_i} \otimes \ket{b_i},
    \label{eq:schmidt_decomp}
\end{equation}
where $n_A = \min\big(\dim(\mathcal{H}_A), \dim(\mathcal{H}_B)\big)$.  A maximally entangled state $\ket{\Psi_{m}} \in \mathcal{H}_A \otimes \mathcal{H}_B$ can be expressed as
\begin{equation}
    \ket{\Psi_{m}} = \frac{1}{\sqrt{n_A}} \sum_{j=0}^{n_A - 1} e^{i\theta_j} \ket{c_j} \otimes \ket{d_j},\; \theta_j \in [0, 2\pi].
    \label{eq:max_ent}
\end{equation}
The Fubini-Study distance between $\ket{\psi}$ and $\ket{\Psi_{m}}$ is
\begin{equation}
    D_{F}(\psi, \Psi_{m}) = \sqrt{1 - |\braket{\psi}{\Psi_{m}}|}.
    \label{eq:fubini}
\end{equation}
Minimizing this distance  is equivalent to maximizing the overlap $|\braket{\psi}{\Psi_{m}}|$:
\begin{equation}
    \max_{\{\ket{\Psi_{m}}\}} |\braket{\psi}{\Psi_{m}}|  = \max_{(\theta_j,\ket{c_j},\ket{d_j})}  \left| \sum_{i,j} \sqrt{\lambda_i} e^{i\theta_j} \braket{a_i}{c_j} \braket{b_i}{d_j} \right|.
\end{equation}
From the triangle inequality, the overlap is maximized when
\begin{equation}
    \theta_j = \text{constant} \; \forall j, \quad 
   \text{ and } \{\ket{c_j}\}, \{\ket{d_j}\} = \{\ket{a_i}\}, \{\ket{b_i}\} .
    \label{eq:max_cond}
\end{equation}
Substituting Eq.~\eqref{eq:max_cond} into Eq.~\eqref{eq:max_ent}, the  maximally entangled state closest  to $\ket{\psi}$ is
\begin{equation}
    \ket{\Psi_{c}} = \frac{1}{\sqrt{n_A}} \sum_{i=0}^{n_A - 1} \ket{a_i} \otimes \ket{b_i}.
\end{equation}
The minimum distance  then depends on the sum of all Schmidt coefficients:
\[  D_{F}(\psi, \Psi_{c}) = \sqrt{1-\frac{1}{\sqrt{n_A}}{\sum_i\sqrt{\lambda_i}}}.\]
We normalize this distance using $N_1$, the distance of any product state to its closest maximally entangled state:
\begin{equation}
    N_1 = \sqrt{1 - \frac{1}{\sqrt{n_A}}}.
    \label{eq:N_1}
\end{equation}
We now express the  SGM  in terms of the Schmidt coefficients as:
\begin{equation}
    \text{SGM}(\psi) = 1 - \frac{1}{N_1}\left(1 - \frac{\sum_i \sqrt{\lambda_i}}{\sqrt{n_A}}\right)^{\frac{1}{2}}.
    \label{eq:mgm_closed}
\end{equation}
It is evident that SGM as expressed in Eq.~\eqref{eq:mgm_closed}, is sensitive to variations in all Schmidt coefficients. This subtle modification ensures that the measure is sharp, enabling it to detect changes in entanglement content more effectively.
Fig:~\ref{LU_ineq} demonstrates that the SGM distinguishes the entanglement content of $\ket{\beta(r)}$, defined in Eq.~\eqref{eq:beta_r}, whereas the GM does not.

\section{SGM Defined Using the Reduced Density Operator Space}
G. Vidal~\cite{Vidal_pure_mixed} pointed out that any entanglement measure defined for pure states is in  one-to-one correspondence with an entanglement measure defined using reduced density operators.
 In this section, we show that the closed-form expression of SGM in Eq.~\eqref{eq:mgm_closed} can equivalently be derived using the Riemannian structure of the reduced density operator space.

The Schmidt coefficients of $\ket{\psi}_{AB} \in H_{A}\otimes H_{B}$ correspond to the eigenvalues of the reduced density operators $\rho_A$ and $\rho_B$. The separable state closest to $\ket{\psi}_{AB}$ is determined by the pure state closest to $\rho_{A/B}$ in either of the reduced spaces $\mathcal{M}_{A/B} \ni \rho_{A/B}$. 

Pure states are the extreme points of the reduced density operator space, while the maximally mixed state is represented as $\rho_{\star A/B} = \frac{\mathbbm{1}}{n_A}$, where  $n_{A} \leq n_B$.

The Sharp Geometric Measure (SGM) corresponds to the distance to the closest maximally entangled state. In the pure state space, identifying the closest maximally entangled state involves optimization, but in the reduced density operator space, all maximally entangled states share the same reduced density operator, $\rho_{\star A}$. Therefore, computing the SGM reduces to measuring the distance between $\rho_A$ and $\rho_{\star A}$.

\subsection*{Choice of Metric on $\mathcal{M}$}
While the search for the closest maximally entangled state simplifies in the reduced density operator space, a metric must still be chosen. In the pure state space, the Fubini-Study metric is the unique Riemannian metric. However, in the mixed state space, multiple Riemannian metrics exist. 

A natural choice for this task is the Bures metric \cite{Bures_orginal_paper}, which coincides with the Fubini-Study metric when restricted to pure states \cite{Braunstein1994}. The Bures metric can be expressed in terms of the fidelity $\mathcal{F}(\rho_1, \rho_2)~=~\Tr\big(\sqrt{\sqrt{\rho_1} \rho_2 \sqrt{\rho_1}}\big)$ as:
\begin{equation}
D^{2}_{\mathcal{B}}(\rho_1, \rho_2) = 2\big(1 - \mathcal{F}(\rho_1, \rho_2)\big).
\label{bures_form}
\end{equation}

The entanglement measure  is actually independent of  the choice of metric. This is seen as follows. In the reduced density operator space for entangled states, the reduced density matrices are LU-equivalent to diagonal matrices (Schmidt form). According to the Morozova-\v{C}entsov-Petz (MCP) Theorem \cite{Morozova1989, Petz1996}, all Riemannian metrics differ only in terms involving the off-diagonal elements of the density matrices. On the submanifold of diagonal density operators, all Riemannian metrics, including the Bures metric, are equivalent to the Fisher-Rao metric \cite{petz_fisher_bures}. Consequently, using the Bures metric for defining the SGM entails no loss of generality.

In terms of distances in the space of reduced density matrices, the Sharp Geometric Measure for a pure state $\ket{\psi}_{AB}$ is
\begin{equation}
    \text{SGM}(\psi_{AB}) = 1 - \frac{D_{\mathcal{B}}(\rho_{A}, \rho_{\star A})}{N_2},
\end{equation}
where $N_2$ is a normalization constant. 

The Bures distance of a pure state $\rho_p$ from $\rho_{\star}$ depends on the dimensionality $\dim[\mathcal{M}] = n$ of the reduced state space. The square-root fidelity evaluates to $\frac{1}{\sqrt{n_A}}$, giving:
\begin{equation}
N_2 = D_{\mathcal{B}}(\rho_{\star}, \rho_p) = \sqrt{2}\left(1 - \frac{1}{\sqrt{n_{A}}}\right)^{\frac{1}{2}}.
\end{equation}
If  $\lambda_i$  are the eigenvalues of  $\rho_A$, the Bures distance from $\rho_{\star A} = \frac{\mathbbm{1}}{n_A}$ is
\begin{equation}
D^2_{\mathcal{B}}(\rho_A, \rho_{\star A}) = 2\left(1 - \frac{1}{\sqrt{n_A}}\sum_i \sqrt{\lambda_i}\right).
\end{equation}
This expression for the SGM is identical to Eq.~\eqref{eq:mgm_closed}.  
Hence the closed-form expression for SGM can be independently derived using either the composite Hilbert space or the reduced density operator space.

\bigskip
\noindent{\bf{Properties of SGM}}

\begin{itemize}
\item \textbf{Local Unitary (LU) invariance:}\\
The form Eq:~\eqref{eq:mgm_closed} ensures that the SGM is LU-invariant since local unitary operators on $\ket{\psi_{AB}}$ will not change the eigenvalues of the reduced density operators. 

\item \textbf{Strong Discriminance:}\\
$N_1$ and $N_2$ are tailored to ensure that for  separable state $\ket{\phi_s}$, 
\begin{equation*}
    SGM(\phi_s) = 0.
\end{equation*} 
Any entangled state $\ket{\Psi_e}$ has more than one Schmidt coefficient which ensures 
\begin{equation*}
SGM(\Psi_e) \neq 0.    
\end{equation*}
\item \textbf{Monotonicity under LOCC:}\\
If the  state $\ket{\psi}_{AB}$ undergoes LOCC and the eigenvalues of the reduced density operator change 
\begin{align*}
       \lambda_i \; \xrightarrow{LOCC} \; \lambda^{\prime}_i,
\end{align*}
 Nielson's majorising theorem \cite{Nielsen_1999} assures us that 
\begin{equation}
    \lambda_{\psi} \succeq \lambda_{\psi^{\prime}}
    \implies \sum_i \sqrt{\lambda_i}  \geq  \sum_i \sqrt{\lambda^{\prime}_i}.
    \label{eq3}
\end{equation}
Using this relation in \eqref{eq:mgm_closed} shows that the SGM is monotonic under LOCC.  
\end{itemize}
Thus, SGM satisfies the desirable properties of a good entanglement measure.

 \section{Genuine Multi-partite Entanglement}\label{GME}
When  more than two parties are involved, there are several inequivalent  ways in which they can be entangled \cite{3qubit_hint,D_r_2000_3qubit}. For instance in case of three qubits there are two SLOCC inequivalent classes of entanglement namely: the GHZ-class  equivalent to 
\begin{equation}
 \ket{GHZ}= \frac{1}{\sqrt{2}} \big(\ket{000} + \ket{111} \big) \label{GHZ}   
\end{equation}
and  the W-class which are superpositions of states in which  only one qubit is in the state $\ket{1}$:
\begin{equation}
    \ket{W}= \frac{1}{\sqrt{3}} \big(\ket{001} + \ket{010} + \ket{100} \big) \label{W}.
\end{equation}
These classes are also generalized to $n$ qubits in a straightforward manner.
For $n>3$, the system could be  in a state that is a tensor product of  entangled subsystems. Genuine entanglement is defined as that for states that are not bi-separable in any way.


We have defined the properties of measure of GME  in the Introduction.
Now it was pointed out in by Love \textit{et al}\cite{Love2007} and Yu \textit{et al}\cite{bipartite-gme_YU2004377} that any bipartite measure can be used to define a measure of GME. We therefore define a measure of GME using the SGM. 

\subsection{GME measure based on SGM}
We define a measure of  GME  in an $n$-partite system as the geometric mean of SGMs of all possible bipartitions.  We call this the GMS (Geometric mean of  Sharp Geometric Measure):
We find the SGM of each bipartition ($b_i$) and calculate the geometric mean of all of them,
\begin{equation}
    \text{GMS}\big(\psi_n\big) = \prod_{i=1}^{m}(b_{i})^{\frac{1}{m}}. \label{eq:GMG}
\end{equation} 

The number $m$  of unique possible bipartitions is given by\cite{GM_of_conc},
\[
m = \left\{ \begin{array}{lr}
        \sum_{i = 1}^{(n-1)/2} {n \choose i} & \text{for odd }n, \\
         &\\
         \sum_{i=1}^{(n-2)/2}  {n \choose i} + \frac{1}{2}{n \choose n/2} & \text{for even }n. 
         \end{array} \right.
\]

\bigskip
\noindent{\bf{Properties of GMS}}

 GMS  satisfies the  properties 
\textbf{(P1)} and \textbf{(P2)}  since it is defined using the  product of bipartite measures.  Based on results in \cite{proper_gme} Xie \textit{et al} defined \textit{proper} GME measure\cite{conc_fill} if the measure ranks entanglement in GHZ state higher than that of W state. Calculations show that GMS$(GHZ) = 1,$ while  GMS$(W) = 0.94$, thus satisfying the condition for \textit{proper} GME measure.

\section{Discussion}

In this section we demonstrate that the  ability to distinguish LU-inequivalent entanglement extends to multipartite entanglement as well.
 
 Based on Generalised Schmidt Decomposition (GSD) there are six LU inequivalent sub-classes\cite{Classification_ent2008, acin_2000_class} for three qubit states.
 Among these are four classes  of genuinely entangled states. They can be identified by the non-zero coefficients in their GSD. The GSD of a 3-qubit pure state takes the canonical form
 \begin{equation*}
    \ket{\psi} =  \lambda_0\ket{000} + \lambda_1\ket{100} + \lambda_2\ket{101} + \lambda_3\ket{110} + \lambda_4\ket{111}.
\end{equation*}
 The four classes each have  non-zero $\lambda_0$ and $\lambda_4$, and are differentiated by the  the non-vanishing of the other $\lambda$s:
\begin{itemize}
    \item[C1:] 
         all of  $\lambda_1, \lambda_2, \lambda_3$ vanish, (GHZ-class);
    \item[C2:] 
         any two of  $\lambda_1, \lambda_2, \lambda_3$ vanish;
    \item[C3:]
        any one of  $\lambda_1, \lambda_2, \lambda_3$ vanishes
    \item[C4:]\label{W-class}:
     none of  $\lambda_1, \lambda_2, \lambda_3$ vanish. 
    (W-class).
\end{itemize}

Several authors have discussed multipartite entanglement measures but few satisfy all the criteria for a measure of GME.

\begin{flushleft}
\begin{figure}
    \centering
\begin{minipage}{\columnwidth}
    \begin{minipage}{0.48\columnwidth}
        \includegraphics[width=\linewidth]{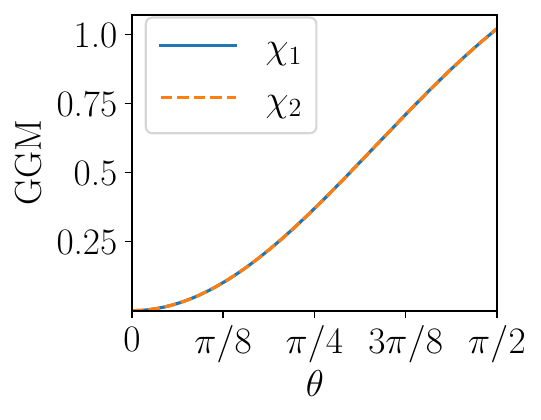}
        \centering {\footnotesize (a) GGM vs $\theta$}
    \end{minipage}%
    \hfill
    \begin{minipage}{0.48\columnwidth}
        \includegraphics[width=\linewidth]{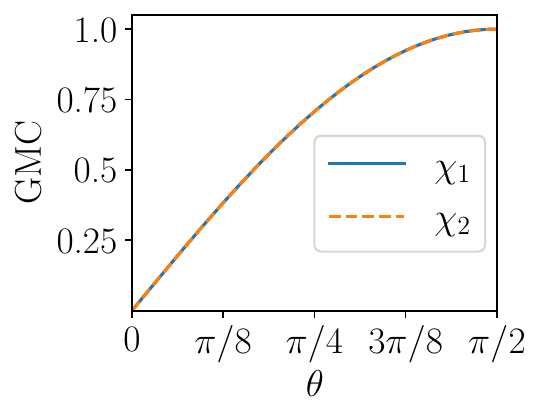}
        \centering {\footnotesize (b) GMC vs $\theta$}
    \end{minipage}
    \vskip 10pt
    \begin{minipage}{0.48\columnwidth}
        \includegraphics[width=\linewidth]{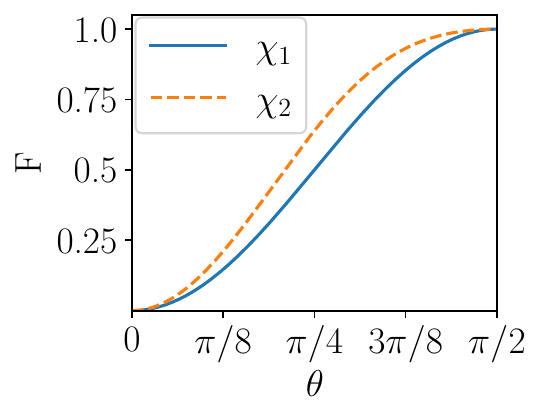}
        \centering {\footnotesize (c) F vs $\theta$}
    \end{minipage}%
    \hfill
    \begin{minipage}{0.48\columnwidth}
        \includegraphics[width=\linewidth]{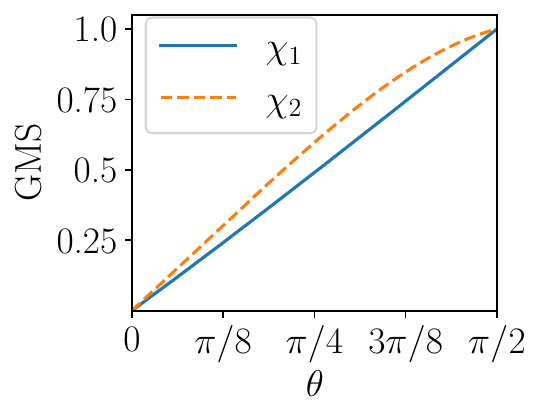}
        \centering {\footnotesize  (d) GMS vs $\theta$}
    \end{minipage}
\end{minipage}
\caption{GME measured for $\chi_1(\theta)$ and $\chi_2(\theta)$ in Eq: \eqref{ex1} using a) GGM, b) GMC, c) F, d) GMS. Concurrence Fill(F) and GMS distinguish the GME of these states belonging to LU-inequivalent classes while GGM and GMC fail to do so.}
    \label{fig:comparison-of-chi}
\end{figure}
\end{flushleft}

Generalised Geometric Mean (GGM)\cite{aditi_sen_HS_dist,ggm_aditi_sen} and Genuinely Multipartite  Concurrence  (GMC)\cite{GMC_eberly} are examples of measures  using the minimum entanglement among all bi-partitions. It turns out that these are relatively poor at  
 discriminating states that belong to the above LU-inequivalent classes. Similar to GM in the bipartite case, these measures depend only  on the largest Schmidt coefficient and hence end up being non-sharp measures of GME.  
 Xie et al\cite{conc_fill} show by  example that Concurrence Fill (F) successfully differentiates  the  entanglement content of  two states belonging to two different sub-classes where GMC and GGM fail. 
 We illustrate that GMS too is successful in discriminating these classes, using two example families of states:
\begin{align}
\text{C1}: \ket{\chi_1(\theta)} &= \cos{\frac{\theta}{2}}\ket{000} + \sin{\frac{\theta}{2}}\ket{111}, \nonumber \\
\text{C2}: \ket{\chi_2(\theta)} &= \frac{1}{\sqrt{2}}\big(\sin{\theta}\ket{000} + \cos{\theta}\ket{110} + \ket{111} \big), \label{ex1}\\
& 0 < \theta \le \frac{\pi}{2}. \nonumber
\end{align}
$\ket{\chi_1(\theta)}$ and $\ket{\chi_2(\theta)}$ belong to classes C1 and C2  respectively. 
Figures~\ref{fig:comparison-of-chi}(a) and (b) show that GMC and GGM put them on equal footing,  whereas F in Figure-~\ref{fig:comparison-of-chi}(c) and GMS in Fig:~\ref{fig:comparison-of-chi}(d) distinguish their entanglement content. 

\begin{figure}[h!]
    \centering
    \includegraphics[width = 0.45\textwidth]{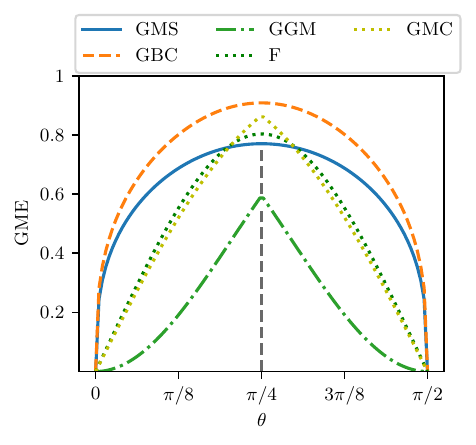}
    \caption{Comparison of GMS with other GME's for the states of class $\chi_3(\theta)$. GMC and GGM show lack of smoothness at $\theta = \pi/4$. }
    \label{ent_measvs_rem}
\end{figure}
\begin{figure}[h!]
   \includegraphics[width=0.42\textwidth]{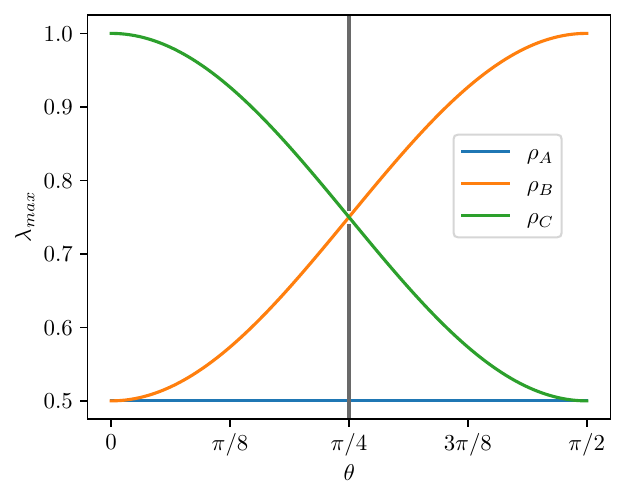} 
\caption{Maximum eigenvalues of three subsystems (A,B,C) of  $\chi_3$. At $\theta = \pi/4$ crossover between maximum eigenvalues of $\rho_B$ and $\rho_C$.}
 \label{schm_dynmcs}
 \end{figure}

Another aspect of  lack of sufficient discriminance of entanglement measures is captured by sharp variations between neighbouring states. 
Consider the variation of entanglement measures with respect to parameter $\theta$ in the family of  states
\begin{equation}
 \ket{\chi_3(\theta)} = \frac{1}{\sqrt{2}}(\sin(\theta)\ket{000} + \ket{011} + \cos(\theta)\ket{110})   
 \end{equation}
As seen in Fig:~\ref{ent_measvs_rem}, 
GGM and GMC show cusps in their variation with $\theta$. 

At first it might appear that this discontinuity in the slope could be an artefact of numerical simulation. However, considering that some of the measures are defined based on the maximum eigenvalue of the reduced density operator,  we trace the source of this behavior to such a non-analytic definition of the   entanglement measures. The plots of maximum eigenvalues of all three subsystems shown in Fig: \ref{schm_dynmcs} show a cross-over fro one subsystem to another at the problem point $\theta = \pi/4$. This is exactly where 
 GGM and GMC show cusps. 
  In contrast, we see that Geometric mean of Bipartite Concurrences (GBC)~\cite{GM_of_conc}, which uses concurrence of bipartitions to measure GME, does not suffer from this problem, neither does our measure, GMS.

GMS compares well with  concurrence fill (F), which is an elegant and visualisable measure for 3-qubit systems. However,  F is hard to generalize to higher number of  qubits. Though there are beautiful entanglement polygon inequalities\cite{Polygon_yang_2022}, there is no reason in general to expect the measure of GME based on bipartite concurrences to behave as area in the $n$-qubit case\cite{trinagle_Guo_2022}.

\section{Conclusion}
In this work, we have defined a bipartite geometric entanglement measure (SGM) based on the distance to the closest maximally entangled state. We also demonstrate that the closed-form expression of SGM can be independently derived in composite bipartite state space as well as the reduced density operator space. In the density operator space, we show that all Riemannian metrics are equivalent for entanglement quantification of pure states.

We have proposed a new feature termed \textit{sharpness} for entanglement measures, which is sensitivity to inequivalence of entanglement under non-local transformations, a property absent in standard Geometric Measures (GMs). The importance of sharpness is highlighted using the teleportation protocol as an illustrative example.  

We further constructed a sharp measure of Genuine Multipartite Entanglement (GME), denoted as GMS, by employing the geometric mean of SGMs. Our measure GMS provides a simple closed-form expression for quantifying GME in any finite-dimensional pure multipartite state. Through examples, we illustrate the advantages of sharp measures of GME, such as GBC and GMS, over other non-sharp measures such as GGM and GMC.

While most resource theories quantify a resource by the distance to the closest non-resource state, we demonstrate that in the case of the entanglement theory of pure states, the distance to the maximal resource state captures more information than the distance to the closest non-resource (separable) state. A similar approach to quantifying coherence was explored by S. Karmakar \textit{et al.}~\cite{coh_fraction}. It is worth exploring whether such a method provides advantages when applied to the quantification of other quantum resources.

\section*{Acknowledgements}
We  acknowledge support from the Department of Science and Technology, Government of India, through the project DST/ICPS/QuST/Theme-3/2019/Q109. We would like to thank Kinjal Banerjee for useful discussions. All simulations were performed using QuTip package \cite{qutip1,qutip2}.

\bibliography{GeometryReferences.bib}
\end{document}